\documentclass[]{piparticle}
\usepackage{graphicx}
\usepackage{amsmath}

\begin{document}

\runningauthor{Authora \itshape{et al.}}  

\title{Random walks on random networks of cliques: \\ Inferring the network structure}

\author{A Nannini,\cite{inst1}\thanks{E-mail: albano.nannini@ib.edu.ar}  
        D H Zanette\cite{inst1,inst2} \thanks{E-mail: damian.zanette@ib.edu.ar}}

\pipabstract{
We study the properties of discrete-time random walks on networks formed by randomly interconnected cliques, namely, random networks of cliques. Our purpose is to derive the parameters that define the network structure --specifically, the distribution of clique size and the abundance of inter-clique links-- from the observation of selected statistical features along the random walk. To this end, we apply a Bayesian approach based on recording the times spent by the walker inside successively visited cliques. The procedure is illustrated with some numerical examples of diverse complexity, where the relevant structural parameters are successfully recovered.   
}

\maketitle

\blfootnote{
\begin{theaffiliation}{99}
   \institution{inst1} Centro Atómico Bariloche and Instituto Balseiro, Comisión Nacional de Energía Atómica and Universidad Nacional de Cuyo, 8400 San Carlos de Bariloche, Río Negro, Argentina.
   \institution{inst2} Consejo Nacional de Investigaciones Científicas y Técnicas, Argentina.
\end{theaffiliation}
}

\section{Introduction}

Formally proposed more than two decades ago \cite{Str1}, network exploration is by now a well-established procedure for disclosing the structure of relational  patterns in a wide variety of complex systems, covering biological and social community detection and identification \cite{commun,bots}, survey of economic (commercial, industrial and financial) complexes \cite{econ}, and big-data mining \cite{
data}, among many other real-life applications. Suitably defined stochastic processes --concretely, random walks endowed with the capacity of recording selected features of their trajectories \cite{internet,random}-- have been pinpointed as a tool to efficiently explore such intricate structures. 

In this contribution, we consider random walks evolving on a specific class of networks, namely, {\em random networks of cliques}. These networks are formed by small groups of fully interconnected nodes --the cliques--  with more sparse, random connections between different groups. They have been introduced as a stylized model of interaction patterns with highly clustered architecture, for which many structural properties can be exactly computed \cite{sob1}. Our main aim is to relate, both analytically and numerically, the statistical properties of the random walk to the structural features of the underlying network, with the ultimate purpose of deriving the latter from the former.     

In the next section, we recall the construction of random networks of cliques and review some of their structural features, which are necessary for our subsequent analysis. In Section \ref{III}, random-walk statistical properties are analytically derived from those structural features. In Section \ref{IV}, we study the inverse problem of deriving the network structure from the observation of the random walk, using a Bayesian inference approach. Three illustrative numerical examples are presented. Finally, we draw our conclusions in Section \ref{V}.

\section{Random networks of cliques (RNoCs)}

\begin{figure}[th]
\begin{center}
\includegraphics[width=.8\columnwidth]{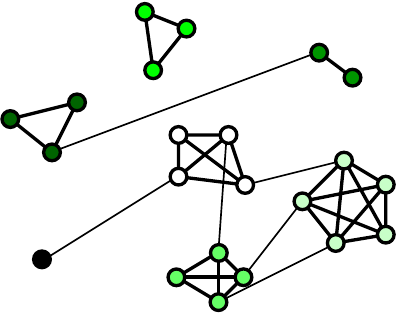}
\end{center}
\caption{A random network of cliques with $Q=7$ cliques, $N=22$ nodes and $M=6$ inter-clique links. The fraction of nodes with inter-clique links is $\gamma \approx 0.55$. In the plot, inter-clique links are thinner than links inside cliques, and nodes in different cliques are colored with different shades.} \label{fig1}
\end{figure}

A random network of cliques (RNoC) is built by first constructing $Q$ cliques. We recall that a clique is a (typically small) group of nodes fully connected to each other. The size $n$ of each clique --namely, its number of nodes--  is drawn at random from a prescribed probability distribution $f_n$ ($n=1,2,\dots$). The expected total number of nodes in the network is thus $N=Q\langle n\rangle$, with $\langle n \rangle$ the mean value of $n$ over the distribution $f_n$. Then, $M$ {\em inter-clique} links are established between randomly chosen nodes belonging to different cliques, with the condition that at most one inter-clique link reaches any given node. The number $M$ is chosen in such a way that the resulting fraction of nodes connected to inter-clique links has a prescribed value $\gamma$. This is achieved by taking $M=N\gamma /2$. Figure \ref{fig1} shows an example of this construction for a small number of cliques.  For future reference, we quote the probability $f_{n,m}$ of having a clique of size $n$ with $m$ inter-clique links $(m\le n)$:
\begin{equation} \label{fnm}
    f_{n,m} = \binom{n}{m} \gamma^n (1-\gamma)^{n-m} f_n.
\end{equation}

As shown elsewhere \cite{sob1}, the simple rules used to build RNoCs enable a straightforward calculation of their structural properties, such as degree distribution, clustering, assortativity, and diameter. In many respects, the RNoC structure is similar to that of small-world networks \cite{sob2,SW}. To our present purposes, a crucial property to assess whether a giant connected component is present or not. In fact, for a random walk to sample a statistically significant part of the network, it must take place on the giant component. We recall that, strictly speaking, a giant component is well defined for an infinitely large network only, as a connected component that contains a finite fraction of the whole network \cite{Newman}. In finite, but large, networks, it is identified as the largest connected component. 

Since cliques are internally fully connected, the existence of a giant component in RNoCs is determined by the pattern of inter-clique links. It can be shown that a giant component comprising a fraction 
\begin{equation} \label{s}
    s=1-G_0(u)
\end{equation}
of the whole network exists when the equation $u=G_1(u)$ has a solution $0<u<1$ \cite{RN}. Here,
\begin{equation} \label{G0}
G_0(x) = \langle (1-\gamma +\gamma x)^n \rangle, 
\end{equation}
and
\begin{equation} \label{G1}
G_1(x) = \frac{G_0'(x)}{G_0'(1)}=\frac{\langle n(1-\gamma +\gamma x)^{n-1 }\rangle}{\langle n \rangle}.
\end{equation}
As above, $\langle \cdot \rangle$ denotes the average over the clique size distribution $f_n$. The solution for $u$ effectively exists when the network is sufficiently connected by inter-clique links, i. e. when $\gamma$ is larger than a certain critical value determined by the distribution $f_n$ \cite{sob1}.

To relate the statistical features of a random walk on the giant component to the structural properties of the network, it is important to take into account that the distribution of clique sizes and inter-clique links is not the same inside the giant component as overall the network. Using standard results on the ``microscopic'' structure of the giant component \cite{ido}, we find that the joint distribution of clique sizes and inter-clique connections inside it is  
\begin{equation} \label{fG}
    f_{n,m}^G = \frac{1-u^m}{s} f_{n,m},
\end{equation}
where $u$ and $s$ are given by Eqs.~(\ref{s}) to (\ref{G1}), and $f_{n,m}$ is the corresponding distribution all over the RNoC, Eq.~(\ref{fnm}).

\section{Statistics of random walks on RNoCs} \label{III}

Now, we consider a random walker moving on the giant component of a RNoC in discrete time. At each time step, the walker jumps from the node it occupies to one of the neighbor nodes, chosen at random with equal probability. The new node can belong to the same clique or, if the original node had an inter-clique link,  to a different one.

Having in mind the aim of inferring the structural features of the RNoC from the random walk statistics, our main interest is to find the frequency with which the walker stays for a certain time in a clique before jumping to a different clique. To this end, we first compute the probability $\Pi_{n,m} (T)$ that the walker spends exactly $T$ steps inside a clique of size $n$ with $m$ inter-clique links. Along a single visit of the random walker to any given clique, the evolution can be conceived as a Markov process with three different states:
\begin{enumerate}
    \item the walker is on a node {\em without} inter-clique link;
    \item the walker is on a node {\em with} inter-clique link;
    \item the walker has left the clique.
\end{enumerate}
The time-dependent probabilities of each state can be arranged in a vector ${\bf p} (t)= [p_1 (t),p_2 (t),p_3 (t)]$, which evolves in time as ${\bf p} (t+1) ={\cal M} {\bf p} (t)$. For a clique with $n$ nodes and $m$ inter-clique links ($1\le m \le n$), the transference matrix reads
\begin{equation} \label{calM}
   {\cal M}= \left( 
    \begin{array}{ccc}
      (n-m-1)/(n-1)   & (n-m)/n & 0  \\
       m/(n-1)  & (m-1)/n &  0 \\
       0 & 1/n & 1
    \end{array}
    \right).
\end{equation}
The probabilities at time $t$, given by
\begin{equation}
  {\bf p}(t)={\cal M}^{t-1} {\bf p} (1) , 
\end{equation}
have to be evaluated using the initial condition ${\bf p}(1)=(0,1,0)$. The elements of $\cal M$ in Eq.~(\ref{calM}) result from straightforward counting of the events that lead from one state to another.

The probability $\Pi_{n,m} (T)$ is given by the product of the probability that the walker is in state $2$ at time $T$, times the probability that it leaves the clique in the next jump, $1/n$. Namely, 
\begin{equation} \label{Pi1}
    \Pi_{n,m} (T)= \frac{1}{n} p_2 (T) = \frac{1}{n} {\bf p} (1) \cdot {\cal M}^{T-1} {\bf p} (1),
\end{equation}
where $\cdot$ indicates scalar product. An explicit form for $\Pi_{n,m} (T)$ can be obtained by diagonalizing  $\cal M$, whose eigenvalues turn out to be $\lambda_1=1$ and
\begin{equation}
    \lambda_{2,3}= \frac{(n-1)^2-m \mp R}{2n(n-1)}.
\end{equation}
Here, $R=\sqrt{(n^2-1)^2-2m(n-1)^2+m^2}$, and the upper and lower signs correspond to $\lambda_2$ and $\lambda_3$, respectively. Operating with the diagonalized version of $\cal M$ in Eq.~(\ref{Pi1}), we find
\begin{align} 
    \Pi_{n,m} (T)= &  \lambda_2^{T-1} \frac{R+n^2-2nm+m-1}{2nR}
   \nonumber \\ & + \lambda_3^{T-1} \frac{  R-n^2+2nm-m+1 }{2nR},
\label{pnm}\end{align}
which is correctly normalized for all $n$ and $m$:
\begin{equation} \label{normPi}
    \sum_{T=1}^\infty \Pi_{n,m} (T)= 1.
\end{equation}
For $n>1$ and $m\le n$, the eigenvalues satisfy $\lambda_2 <0<\lambda_3$ and $|\lambda_2|<\lambda_3 < 1$. These inequalities imply that, for large $T$, $\Pi_{n,m} (T)$  is dominated by the second term in Eq.~(\ref{pnm}), and therefore decreases as $\Pi_{n,m} (T) \sim \lambda_3^T$. Moreover, since $\lambda_2$ is negative, an oscillatory dependence is expected for small $T$ due to the alternating sign of the first term. 

Once $\Pi_{n,m} (T)$ has been evaluated, we can compute the probability $P(T)$ that a random walker on the giant component of a given RNoC stays exactly $T$ steps inside any of its cliques.  This probability results from a sum of the contributions of cliques of each size $n$ and each number of inter-clique links $m$ weighted by the frequency of such cliques, which is given by the probability distribution $f_{n,m}^G$ obtained in the preceding section, Eq.~(\ref{fG}). Moreover, each contribution is weighted by the probability of entering a given clique, which is proportional to the number of its inter-clique links, $m$. The result is
\begin{equation} \label{P}
    P(T)=\frac{\sum_n \sum_{m=1}^n m f_{n,m}^G \Pi_{n,m} (T)}{\sum_n \sum_{m=1}^n m f_{n,m}^G},
\end{equation}
with the first sum running over the relevant values of $n$, which depends on the distribution of clique sizes. The normalization of $P(T)$ is insured by that of $\Pi_{n,m} (T)$, Eq.~(\ref{normPi}). Generally, $P(T)$ cannot be given an explicit analytical form but can be straightforwardly computed by numerical means once $f_{n,m}^G$ has been specified.

Direct application of the result for $P(T)$, Eq.~(\ref{P}), to the statistics of the time spent inside each clique along an actual random walk on a RNoC requires neglecting correlations between successive visits to different cliques. In fact, after abandoning a given clique to enter a neighbor, there is a relatively large probability that the walker returns to the previous clique. This kind of correlation, however, should become less important as the random walk progresses, covering increasingly large portions of the network. 

\begin{figure}
\begin{center}
\includegraphics[width=\columnwidth]{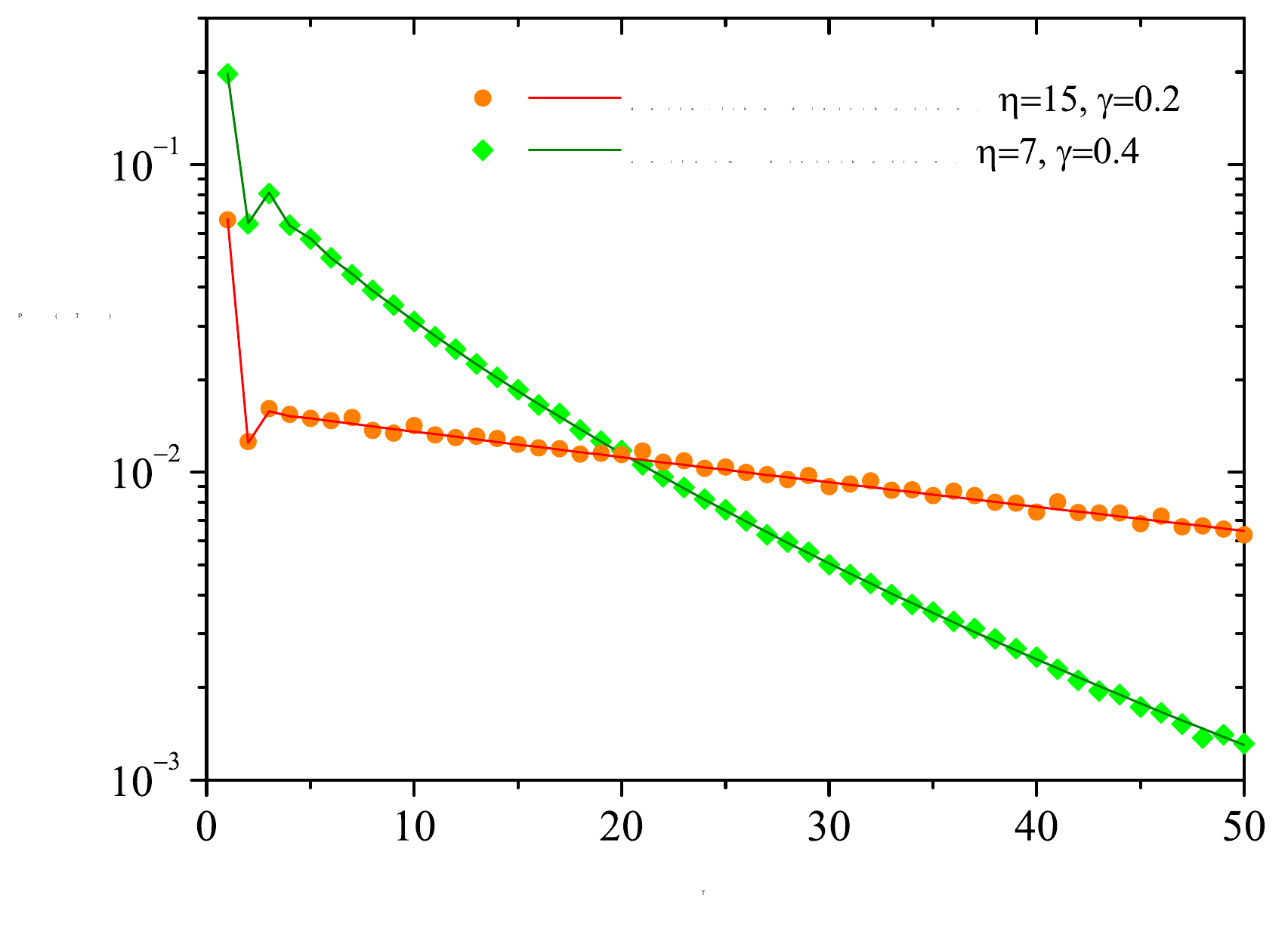}
\end{center}
\caption{Probability $P(T)$ that the random walker spends $T$ steps inside a clique, Eq.~(\ref{P}), for two distributions of clique sizes $f_n$, Eqs.~(\ref{fdelta}) and (\ref{funif}), and two values of the fraction of nodes with inter-clique links, $\gamma$. Symbols correspond to simulation results, while lines join the values of $P(T)$ computed using Eq.~(\ref{P}).} \label{fig2}
\end{figure}

To confirm this conjecture, we have simulated random walks on the giant component of RNoCs with two distributions of clique sizes, namely, a delta-like distribution,
\begin{equation} \label{fdelta}
    f_{n} = \delta_{n,\eta},  
\end{equation}
where all cliques have size $\eta$, and a uniform distribution,
\begin{equation} \label{funif}
  f_n =  \begin{cases}
 1/(\eta-2) & \mbox{ for $3\le n\le \eta$},\\
 0 & \mbox{ otherwise},
\end{cases}
\end{equation}
where all the sizes between $3$ and $\eta$ have the same probability. Depending on the value of $\eta$, the fraction of nodes with inter-clique links, $\gamma$, has been chosen to ensure the presence of a well-developed giant component. Each random walk was conducted on a RNoC consisting of $Q=10^5$ cliques, along $10^7$ time steps. Figure \ref{fig2} shows the results in two illustrative cases: the delta-like distribution, Eq.~(\ref{fdelta}) with $\eta=15$, for $\gamma=0.2$, and the uniform distribution, Eq.~(\ref{funif}) with $\eta=7$, for $\gamma=0.4$. Symbols stand for the results of numerical realizations of the random walk, and lines join the values of $P(T)$ obtained from Eq.~(\ref{P}) as a function of $T$. Up to small random fluctuations, the excellent agreement between numerical and analytical results strongly supports the above conjecture. Moreover, the plots illustrate the oscillations expected in $P(T)$ for small values of $T$ due to the opposite signs of $\lambda_2$ and $\lambda_3$ in Eq.~(\ref{pnm}), as well as the exponential decrease for large $T$.

\section{Bayesian inference of the RNoC structure} \label{IV}

To apply the above results to the inference of the RNoC structure, we assume that, as the result of observing a random walk on the network, a vector ${\bf T}= ( T_1, T_2, \dots, T_K)$ has been obtained, whose components $T_k$ are the times spent by the walker inside successively visited cliques. Our aim is to compute the conditional probability $P({\bf \Phi}| {\bf T})$ that the network has a certain structure given the observed vector $\bf T$. Here, $\bf \Phi$ denotes a vector whose components are the parameters that specify the RNoC structure. These encompass the fraction of nodes with inter-clique links, $\gamma$, and the parameters needed to determine the distribution of clique sizes $f_n$. Generally, among the components of $\bf \Phi$, some parameters may be known with certainty, while others are to be determined from probabilistic inference.  

By virtue of Bayes' theorem \cite{Bayes}, we have
\begin{equation}  \label{Bayes}
    P({\bf \Phi}| {\bf T}) = \frac{P( {\bf T}| {\bf \Phi}) P({\bf \Phi})}{P( {\bf T})} ,
\end{equation}
where $P( {\bf T}| {\bf \Phi})$ is the probability of observing the vector $\bf T$ conditioned to the parameter set $\bf \Phi$, $P({\bf \Phi})$ is the {\em a priori} probability assigned to the set $\bf \Phi$, and  $P( {\bf T}) = \sum_{{\bf \Phi}} P( {\bf T}| {\bf \Phi}) P({\bf \Phi})$. The conditional probability $P( {\bf T}| {\bf \Phi})$ can be computed using the results obtained in Section \ref{III}, as
\begin{equation} \label{Pcond}
    P( {\bf T}| {\bf \Phi}) = \prod_{k=1}^K
P(T_k| {\bf \Phi}),
\end{equation}
where $P(T_k| {\bf \Phi})$ is the probability $P(T)$ given by Eq.~(\ref{P}), calculated for an RNoC with parameters $\bf \Phi$ and evaluated on time $T_k$. In Eq.~(\ref{Pcond}), we have assumed that the probabilities of the components of $\bf T$ are mutually independent, as supported by the numerical results presented at the end of the preceding section. 

In order to minimize the information provided to evaluate $P({\bf \Phi}| {\bf T})$, we assign the same {\em a priori} probability to all the sets $\bf \Phi$. With this choice, Eq.~(\ref{Bayes}) can be rewritten as
\begin{equation}  \label{B2}
    P({\bf \Phi}| {\bf T}) = \frac{P( {\bf T}| {\bf \Phi}) }{Z( {\bf T})} ,
\end{equation}
with $Z( {\bf T})= \sum_{{\bf \Phi}} P( {\bf T}| {\bf \Phi})$. Using Eq.~(\ref{B2}), in practice, we compute $P({\bf \Phi}| {\bf T})$ by first evaluating $P( {\bf T}| {\bf \Phi})$ from Eq.~(\ref{Pcond}), and then normalizing over the relevant values of $\bf \Phi$. Once $P({\bf \Phi}| {\bf T})$ has been obtained as a function of $\bf \Phi$, the unknown network parameters inferred from the vector $\bf T$ are those with the highest probability, namely those that maximize $P({\bf \Phi}| {\bf T})$.

To assess the performance of this procedure, we operate as follows. First, we generate a RNoC with a prescribed set of parameters ${\bf \Phi}_0$. Next, a random walk is carried out on its giant component, recording the times $T_k$ spent by the walker in the successive cliques it visits. This provides the vector ${\bf T}= ( T_1, T_2, \dots, T_K)$. Then, using Eqs.~(\ref{Pcond}) and (\ref{B2}), we calculate the conditional probability $P({\bf \Phi}| {\bf T})$  and detect the parameter set ${\bf \Phi}_{\max}$ for which this probability reaches a maximum. Finally, ${\bf \Phi}_{\max}$ is compared with ${\bf \Phi}_0$. The larger the coincidence, the better the performance.

\subsection{Inferring $\gamma$}

As a first example, we assume that we know the distribution of clique sizes in the RNoC, $f_n$, but that the fraction of nodes with inter-clique links $\gamma$ is unknown and has to be inferred as explained above. To test the procedure in this situation, we have taken $f_n$ as the delta-like distribution of Eq.~(\ref{fdelta}), with $\eta=7$, and have built a RNoC with $Q=10^5$ cliques and $\gamma_0=0.7$. On this network, we performed random walks of different lengths $t_{\max}$ and, for each random walk, we determined the vector $\bf T$ containing the times spent by the walker inside successively visited cliques. 

\begin{figure}[th]
\begin{center}
\includegraphics[width=\columnwidth]{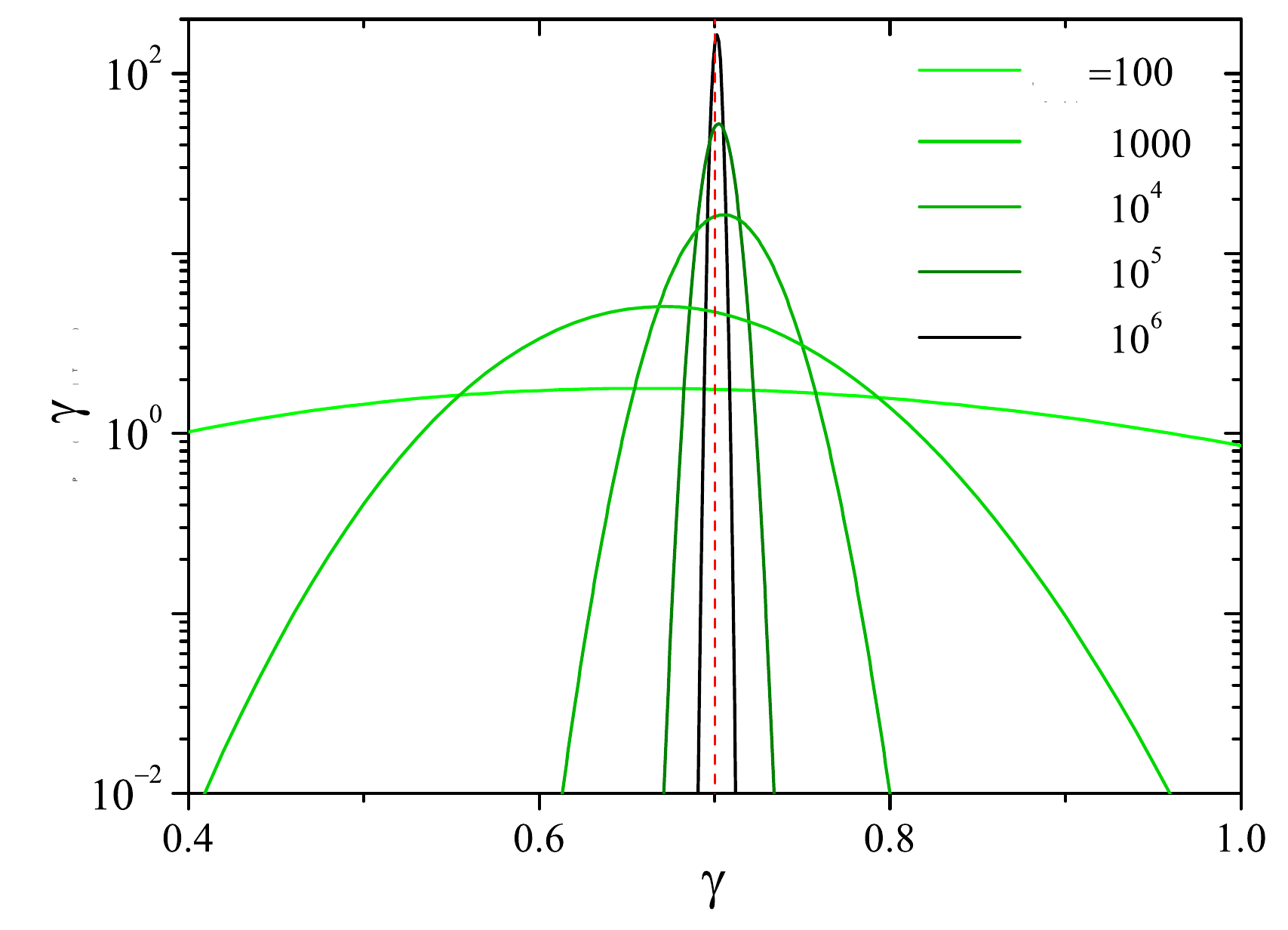}
\end{center}
\caption{The probability distribution $P(\gamma | {\bf T})$ as a function of $\gamma$, computed from six random walks with different length $t_{\max}$ on the giant component of a RNoC with cliques of identical sizes $n=7$. The dashed vertical line indicates the fraction of nodes with inter-clique links used to build the network,  $\gamma_0=0.7$.} \label{fig3}
\end{figure}

Figure \ref{fig3} shows the probability distributions $P(\gamma | {\bf T})$ calculated as prescribed by Eqs. (\ref{Pcond}) and (\ref{B2}), for six random walks with various values of $t_{\max}$. For each walk, $P(\gamma | {\bf T})$ has been computed as a function of $\gamma$ with discretization $\delta \gamma= 10^{-4}$. The results clearly show that $P(\gamma | {\bf T})$ becomes increasingly concentrated around a well-defined value of $\gamma$  as $t_{\max}$ grows and the random walk samples the network more thoroughly.  For $t_{\max}=10^6$, $P(\gamma | {\bf T})$ attains its maximum at $\gamma_{\max}=0.7013(1)$, which constitutes our best estimation for $\gamma$ from this specific realization. The small difference between $\gamma_{\max}$ and $\gamma_0$, just below $0.2$ \%, can be ascribed to the fact that inter-clique links in the RNoC are established at random as it is built, so that the fraction of nodes with such links in a given realization of the network can slightly differ from $\gamma_0$.    

\begin{figure}[th]
\begin{center}
\includegraphics[width=\columnwidth]{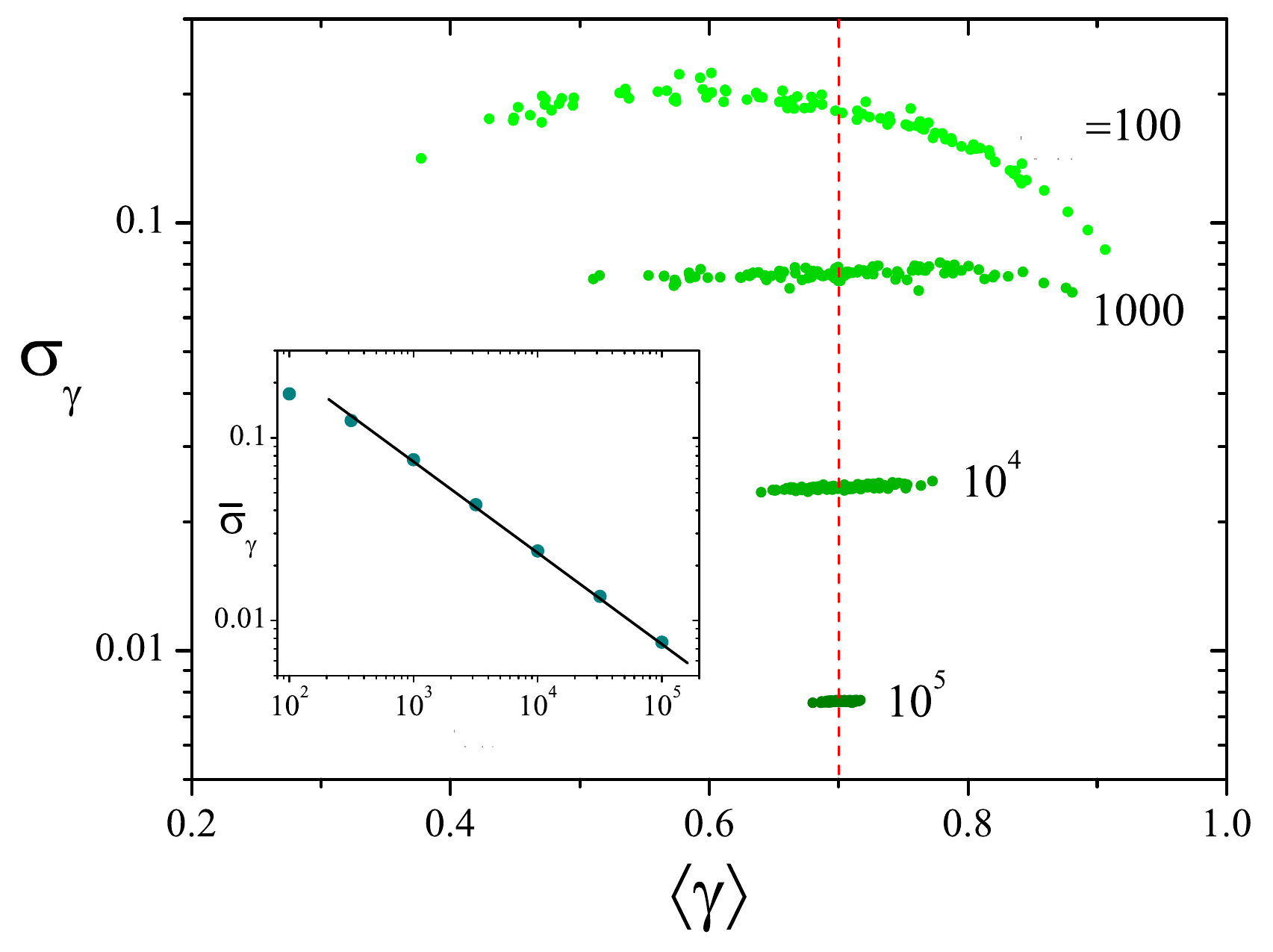}
\end{center}
\caption{Main panel: Mean value $\langle \gamma \rangle$ and standard deviation $\sigma_\gamma$ for the probability distribution $P(\gamma | {\bf T})$ obtained from random walks of length $t_{\max}$ on a RNoC with cliques of identical sizes $n=7$. Each dot corresponds to a single random walk, among $100$ for each value of $t_{\max}$. The dashed vertical line indicates the fraction of nodes with inter-clique links used to build the network,  $\gamma_0=0.7$.  Inset: The standard deviation averaged over realizations, $\bar \sigma_\gamma$, as a function of $t_{\max}$. The slope of the straight segment equals $-1/2$.} \label{fig4}
\end{figure}

To evaluate the dispersion in the results obtained from different realizations of the random walk, we performed sets of $100$ random walks for each value of $t_{\max}$. For each random walk, we have computed  $P(\gamma | {\bf T})$ as a function of $\gamma$ --now, with discretization $\delta \gamma= 10^{-2}$-- and numerically  evaluated the mean value
\begin{equation}
    \langle \gamma \rangle = \int_0^1 \gamma P(\gamma | {\bf T}) d\gamma,
\end{equation}
and the corresponding standard deviation $\sigma_\gamma$. For the peaked profiles of $P(\gamma | {\bf T})$ as in Fig.~\ref{fig3}, the mean value $\langle \gamma \rangle$ yields an excellent estimation of the position of the maximum, $\gamma_{\max}$.  Results are shown in the main panel of Fig.~\ref{fig4}, where each dot corresponds to a single realization. We see that as $t_{\max}$ grows, not only does $\sigma_\gamma$ decrease --as already shown in Fig.~\ref{fig3}-- but the values of $\langle \gamma \rangle$ become more consistent between realizations. The inset shows the standard deviation averaged over realizations, $\bar \sigma_\gamma$, as a function of $t_{\max}$. The straight segment has a slope of $-1/2$, indicating that the average standard deviation decreases as $\bar\sigma_\gamma \sim t_{\max}^{-1/2}$ along the considered interval. 

We mention that we have also tested the complementary task of inferring $\eta$ for delta-like and uniform distributions, Eqs.~(\ref{fdelta}) and (\ref{funif}), now assuming that $\gamma$ is known. In this case, successful inference requires shorter random walks, because $\eta$ is a discrete (integer) parameter. In fact, the probability $P(\eta | {\bf T})$ already has a maximum at the correct value of $\eta$ for random walks of length between $10^3$ and $10^4$. For brevity, the results for these cases are not presented here.

\subsection{Simultaneous inference of $\gamma$ and $\eta$}

In our second example, we use the above procedure to infer two parameters, namely, $\gamma$ and $\eta$, assuming that the clique sizes are uniformly distributed according to Eq.~(\ref{funif}). To test the method, we constructed a RNoC with $10^5$ cliques using $\gamma_0=0.75$ and $\eta_0=10$. The results presented here were obtained from three realizations of the random walk on the giant component of the network, with $t_{\max}=10^4$, $10^5$, and $10^6$ steps, respectively. 

\begin{figure}[th]
\begin{center}
\includegraphics[width=\columnwidth]{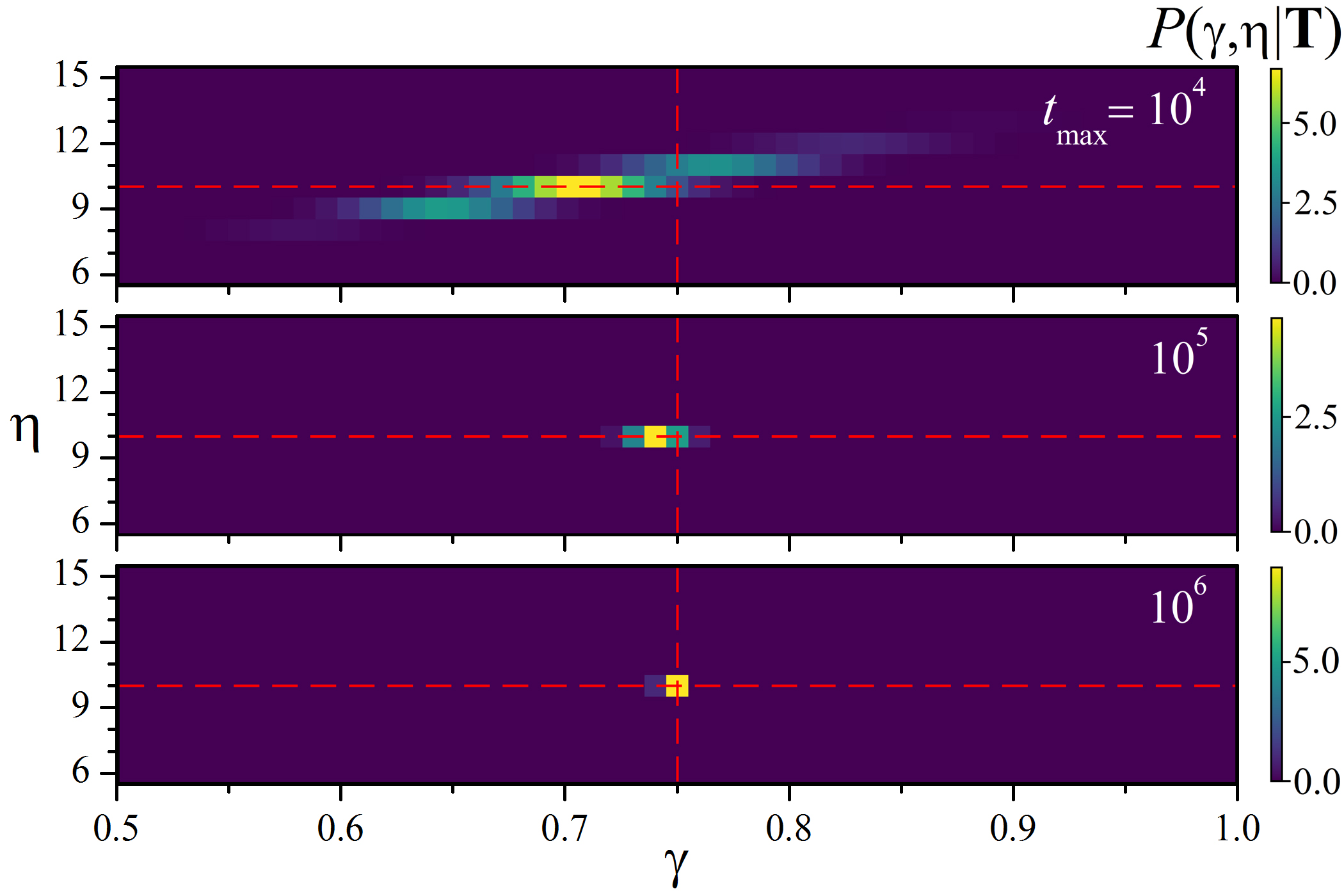}
\end{center}
\caption{The probability $P(\gamma,\eta|{\bf T})$, in a color scale, on the plane spanned by the parameters $\gamma$ and $\eta$, as estimated from three random walks of different length $t_{\max}$ on the giant component of a RNoC with $\gamma_0=0.75$ and a uniform distribution of clique sizes, Eq.~(\ref{funif}), with $\eta_0=10$. The discretization step in $\gamma$ is $\delta\gamma=10^{-2}$. Straight dashed lines stand for the values of $\gamma_0$ and $\eta_0$. } \label{fig5}
\end{figure}

Figure \ref{fig5} depicts, in a color scale, the probability distribution $P(\gamma,\eta|{\bf T})$ on the plane $(\gamma,\eta)$, as obtained from Eqs. (\ref{Pcond}) and (\ref{B2}) using the vectors $\bf T$ recorded along the three random walks. Dashed lines stand for the values of $\gamma_0$ and $\eta_0$. The plots show clearly that, much as observed in the previous example, the probability becomes more concentrated and better centered around the expected values of the parameters as the random walks increase in length. 

Table \ref{table1} presents the mean values and standard deviations of $\gamma$ and $\eta$ derived from the distribution $P(\gamma,\eta|{\bf T})$ for each random walk. Note that the estimate of the discrete parameter $\eta$ converges faster and more accurately to its expected value than that of $\gamma$. In any case, for $t_{\max}=10^6$, both $\gamma_0$ and $\eta_0$ lie inside the intervals defined by the mean value and the standard deviation of each parameter.

\begin{table}[h]
\centering
\begin{tabular}{c c c c c} 
 \hline
$t_{\max}$ & $\langle \gamma \rangle$ & $\sigma_\gamma$ & $\langle \eta \rangle$ &  $\sigma_\eta$ \\  
 \hline\hline 
 $10^4$ & $0.71$ & $0.06$ & $10.1$ & $0.9$ \\ 
 $10^5$ & $0.741$ & $0.009$ & $10.00$ & $0.04$ \\  
 $10^6$ & $ 0.749$ & $0.003 $ & $10.00$  & $< 10^{-5}$ \\
 \hline
\end{tabular}
\caption{Mean value and standard deviation of the parameters $\gamma$ and $\eta$ deriving from the estimate of the probability $P(\gamma,\eta|{\bf T})$, as obtained from the three random walks described in the text.}
\label{table1}
\end{table}

\subsection{Parameter inference by gradient ascent}

When the network parameters to be inferred from a random walk on a RNoC are continuous, and/or when their number increases, the calculation of the probability $P({\bf \Phi}| {\bf T})$ all over the multidimensional space spanned by $\bf \Phi$ may become computationally impractical. In such a case, a better strategy may be needed to efficiently locate the point where $P({\bf \Phi}| {\bf T})$ attains its maximum. An alternative procedure consists of implementing a biased random walk on parameter space (not to be confused with the random walk on the RNoC, used to obtain $\bf T$), whose steps only occur in directions along which the probability grows. In this way, $P({\bf \Phi}| {\bf T})$ only needs to be calculated along the random walk path and its immediate neighborhood. This is nothing but a {\em gradient ascent} procedure \cite{optim}, which ensures that, for sufficiently long times, a local maximum of the probability is eventually reached.      

We have implemented this method to infer two continuous parameters of a RNoC whose cliques have two possible sizes ($n=3$ or $4$) distributed with different frequencies, namely,
\begin{equation} \label{fxi}
  f_n =  \begin{cases}
 \xi & \mbox{ for $n=3$},\\
 1-\xi & \mbox{ for $n=4$}, \\
 0 & \mbox{ otherwise},
\end{cases}
\end{equation}
with $0<\xi<1$. The parameters to be estimated are $\gamma$ and $\xi$. In the specific example presented below, we have built the RNoC of $10^5$ cliques using $\gamma_0=0.75$ and $\xi_0=0.5$. Then, we have run a single random walk of $10^7$ steps on its giant component, in order to obtain the vector $\bf T$.  

\begin{figure}[th]
\begin{center}
\includegraphics[width=\columnwidth]{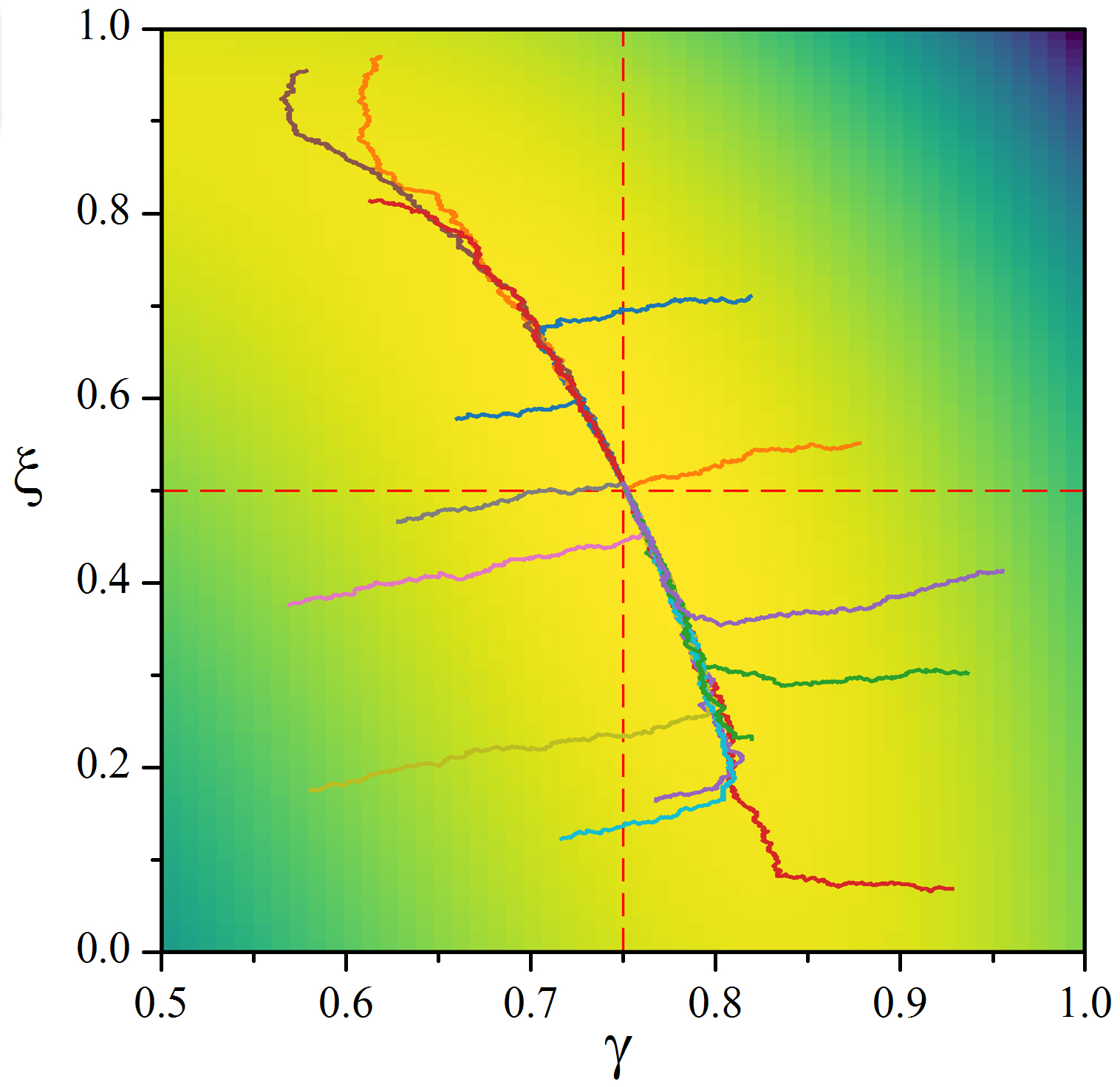}
\end{center}
\caption{Fifteen trajectories of a gradient-ascent random walk on the parameter plane $(\gamma,\xi)$, used to find the maximum of the probability  $P(\gamma,\xi|{\bf T})$ as explained in the text. Dashed lines stand for the values $\gamma_0$ and $\xi_0$ of the RNoC whose parameters are inferred. The background represents $P(\gamma,\xi|{\bf T})$ in a color scale, with lighter shades corresponding to higher values.} \label{fig6}
\end{figure}

The gradient-ascent random walk on the parameter space $(\gamma,\xi)$ --which, for computational representation, is tessellated with discretization $\delta \gamma$ and $\delta \xi$ in each direction-- is carried on as follows. An initial position for the walker, say $(\gamma_i,\xi_i)$, is chosen at random in the relevant parameter interval and the probability $P(\gamma_i,\xi_i|{\bf T})$ is computed. Then, a jump to a randomly chosen nearest-neighbor cell, say  $(\gamma_j,\xi_j)$, is attempted. If $P(\gamma_i,\xi_i|{\bf T})<P(\gamma_j,\xi_j|{\bf T})$ the jump is accepted and the walker moves. Otherwise, another potential new position  $(\gamma_j,\xi_j)$ is selected. If the jump is rejected a prescribed number of times $\tau$, the process ends. This procedure is expected to lead to the ``trapping'' of the random walker in a cell where the probability attains a local maximum.

Figure \ref{fig6} shows $15$ trajectories of the random walker on the plane $(\gamma,\xi)$, starting from different initial positions, with $\delta \gamma=\delta \xi=10^{-3}$ and $\tau=10$. It is apparent that, in all cases, the walker converges toward the expected point $(\gamma_0,\xi_0)$, at the intersection of the dashed lines. The final positions of all the trajectories yield $\gamma=0.750(1)$ and $\xi=0.50(1)$ for the inferred parameters. Note also that all trajectories exhibit a first stage where the walker approaches a curved manifold and then proceed along it. The slope of the manifold near $(\gamma_0,\xi_0)$ indicates that the convergence is considerably faster along $\gamma$ than along $\xi$.

\section{Conclusions} \label{V}

In this paper, we have tested a procedure to derive information on network structure from observing random walks evolving on the network --or, more precisely, on its giant component. Concretely, we have implemented the method for random networks of cliques (RNoCs), for which we derived the parameters defining the distribution of clique sizes and the density of inter-clique connections from the sequence of times spent by the random walker inside successively visited cliques. Parameter estimation was based on Bayesian inference, under the assumption of uniform {\em a priori} probability. Within this formulation, Bayes' theorem allows for the calculation of a probability distribution over parameter space, whose maximum is associated with the best estimation. For two of the examples presented here, we have shown how the results improve in precision as the random walk becomes longer. In all cases, we have obtained excellent estimations.

As we have demonstrated through specific examples, the procedure can be implemented in two alternative ways. When the number of parameters to be inferred is small, the probability distribution can be computed over the relevant zone of parameter space, and its maximum can be detected with satisfactory precision. As the number of parameters grows, however, the calculation on a multi-dimensional space may become computationally very expensive. In this case, the position of the maximum can be found more efficiently using other methods. Here, we have used a random walk driven by gradient-ascent but, if a complicated probability landscape is foreseen, a Monte Carlo algorithm could provide a more exhaustive exploration of parameter space.         

Network exploration using random walks requires that, besides the rules that define its dynamics, the walker (or its observer) is endowed with the capability of recording some selected network features as the process progresses. In our case, we have assumed that it was possible to detect when the walker jumps between cliques, for instance, discerning between intra- and inter-clique connections.  This assumption was essential to provide the series of times spent inside cliques, which fed the Bayesian approach. Inference based on recording other features along the random walk may be the subject of future work.


\end{document}